\documentclass{article}

\newcommand{\mycite}[1]{ \cite{#1}}
\newcommand{\myccite}[2]{ \cite[#1]{#2}}
\usepackage{amsmath,amssymb,a4wide,amsthm}
\usepackage{epsfig}
  
\numberwithin{equation}{section}
 
\newtheorem{theorem}{Theorem}
\newtheorem{prop}{Proposition}
\newtheorem{ass}{Assumption}

\newcommand{\Bild}[4]{ 
\begin{figure}[htb] 
  \begin{center} 
    \leavevmode 
    \epsfig{file=#2,height=#1cm} 
    \caption{{\small #3}} 
    \label{#4} 
  \end{center} 
\end{figure}} 
 
\title{Poland-Scheraga models and the\\ 
DNA denaturation transition} 
 
\author{Christoph Richard\\ 
Institut f\"ur Mathematik und Informatik\\ 
Universit\"at Greifswald, Jahnstr.~15a\\ 
17487 Greifswald, Germany\\ 
\\ 
Anthony J.~Guttmann\\ 
Department of Mathematics and Statistics\\ 
University of Melbourne, Victoria 3010, Australia}

\begin{document} 
 
\maketitle

\begin{abstract} 
Poland-Scheraga models were introduced to describe the DNA denaturation 
transition.  We give a rigorous and refined discussion of a family of these 
models.  We derive possible scaling functions in the neighborhood of the 
phase transition point and review common examples.  We introduce a 
self-avoiding Poland-Scheraga model displaying a first order phase transition 
in two and three dimensions.  We also discuss exactly solvable directed 
examples.  This complements recent suggestions as to how the Poland-Scheraga 
class might be extended in order to display a first order transition, which is 
observed experimentally. 
\end{abstract}

\section{Introduction} 
 
When a solution of DNA is heated, the double stranded molecules denature into 
single strands.  In this process, looping out of AT rich regions of the DNA 
segments first occurs, followed eventually by separation of the two strands as 
the paired segments denature.  This {\em denaturation} process corresponds to 
a phase transition\mycite{WB85}. 
 
A simple model of the DNA denaturation transition was introduced in 1966 by 
Poland and Scheraga\mycite{PS66a, PS66b} (hereinafter referred to as PS) and 
refined by Fisher\mycite{F66, F84}.  The model consists of an alternating 
sequence (chain) of straight paths and loops, which idealize denaturing DNA, 
consisting of a sequence of double stranded and single stranded molecules.  An 
attractive energy is associated with paths.  Interactions between different 
parts of a chain and, more generally, all details regarding real DNA such as 
chemical composition, stiffness or torsion, are ignored.  It was found that 
the phase transition is determined by the critical exponent $c$ of the 
underlying loop class.  Due to the tractability of the problem of random 
loops, that version of the problem was initially studied by PS\mycite{PS66b}. 
The model displays a continuous phase transition in both two and three 
dimensions.  It was argued\mycite{F66} that replacing random loops by 
self-avoiding loops, suggested as a more realistic representation accounting 
for excluded volume effects within each loop, sharpens the transition, but 
does not change its order. 
 
However, the sharp jumps observed in the UV absorption rate in DNA melting%
\footnote{ 
In recent years, a number of other properties of DNA molecules have been 
studied by refined techniques such as optical tweezers and atomic force 
microscopy, and theoretical descriptions of underlying effects such as 
unzipping\mycite{KMP01, MBM01, KMP02, BCS02, MBM02} have been proposed.  These 
will not be discussed here.} 
expe\-riments\mycite{WB85}, which correspond to a sudden breaking of large 
numbers of base pairs, indicated that a first order phase transition would be 
the appropriate description.  The question whether such an asymptotic 
description, which implies very long chains, is valid for relatively short DNA 
sequences, has been discussed recently\mycite{HM01,KMP01b}.  Nevertheless, a 
directed extension of the PS model, being essentially a one-dimensional Ising 
model with statistical weighting factors for internal loops, is widely used 
today and yields good coincidence of simulated melting curves with 
experimental curves for known DNA sequences\mycite{WB85,B+99, BC02}.  Another 
recent application of PS models analyses the role of mismatches in DNA 
denaturation\mycite{GO03}.  A numerical approach to DNA denaturation, which we 
will not discuss further, uses variants of the Peyrard-Bishop 
model\mycite{PB89,TDP00}, a Hamiltonian model of two harmonic chains coupled 
by a Morse potential. 
 
With the advent of efficient computers, it has more recently been possible to 
simulate analytically intractable models extending the PS class, which are 
assumed to be more realistic representations of the biological problem.  One of these is a 
model of two self-avoiding and mutually avoiding walks, with an attractive 
interaction between different walks at corresponding positions in each 
walk\mycite{CCG00, COS02, BCS02, BCKMOS02}.  The model exhibits a first order 
phase transition in $d=2$ and $d=3$.  The critical properties of the model are 
described by an exponent $c'$ related to the loop length distribution\mycite{KMP00, 
  KMP01, KMP02, COS02, BCS02, BCKMOS02}, see also Fisher's review 
article\mycite{F84}.  This exponent is called $c$ again.  Indeed, for PS 
models, it coincides with the loop class exponent $c$ if $1<c<2$, see below. 
Within a refined model, where different binding energies for base pairs and 
stiffness are taken into account, the exponent $c'$ seems to be largely 
independent of the specific DNA sequence and of the stiffness of paired walk 
segments corresponding to double stranded DNA parts\mycite{COS02}.  There are, 
however, no simulations of melting curves for known DNA sequences which are 
compared to experimental curves for this model. 
 
An approximate analytic derivation of the exponent related to this new model 
was given by Kafri et al.\mycite{KMP00, KMP01, KMP02} using the theory of 
polymer networks.  They estimated the excluded volume effect arising from the 
interaction between a single loop and two attached walks.  This approach 
(refined recently\mycite{BCS02}) yields an approximation to the loop length 
distribution exponent $c'$, which agrees well with simulation results of 
interacting self-avoiding walk pairs\mycite{COS02, BCS02, BCKMOS02}.  There is 
a recent debate about the relevance of this approximation to real 
DNA\mycite{HM01,KMP01b}.  The polymer network approach, as initiated 
previously\mycite{KMP00}, led to a number of related 
applications\mycite{KMP01, KMP02, BCS02, BCOS02, HM02}. 
 
In this article, we reconsider PS models for three principal reasons. 
Firstly, the older articles are short in motivating the use of particular loop 
classes, which may be misleading in drawing conclusions about the 
thermodynamic effects of different loop classes.  In fact, the loop classes 
discussed in the early approaches\mycite{PS66b,F66} are classes of {\it 
  rooted} loops and lead to chains which are not self-avoiding.  This seems 
unsatisfactory from a biological point of view, since real DNA is 
self-avoiding.  Secondly, the common view holds that PS models with 
self-avoiding loops cannot display a first order transition in two or three 
dimensions.  In fact, this view led to extending the PS class\mycite{CCG00, 
  KMP00, KMP01, KMP02} in order to find a model with a first order transition. 
However this view is incorrect, as we demonstrate by a self-avoiding PS model 
with self-avoiding loops.  Thirdly, the two exponents $c$ and $c'$, extracted 
from different expressions as described above, are used in the literature 
without distinction, although there are subtle differences, which we will 
point out. 
 
This paper is organized as follows.  In the next section, we give a rigorous 
discussion of PS models and their phase transitions.  We derive the scaling 
functions which describe the behavior below the critical temperature, as the 
critical point is approached.  We then derive the loop statistics, thereby 
analyzing the occurrence of the loop class exponent in the loop length 
distribution.  The third section reviews the prevailing PS models, with 
emphasis on motivation for the underlying loop classes.  We then introduce a 
self-avoiding PS model displaying a first order phase transition in two and 
three dimensions.  In Section \ref{sec:exact}, exactly solvable directed PS 
models are discussed.  We will give explicit expressions for generating 
functions.  The critical behavior of some of these models has been analyzed 
previously by different methods\mycite{L91, KS92, MBM01, MBM02}. This is 
followed by a discussion of models extending the PS class. We conclude with a 
discussion of some open questions. 
 
\section{Poland-Scheraga models: general formalism} 
 
We define PS models and consider analytical properties of their free energies, 
thereby refining previous expositions\mycite{PS66a, PS66b, F66, F84} in a 
rigorous manner.  We analyze phase transitions, extract the asymptotic 
behavior of chains at temperatures near the phase transition by means of 
scaling functions, and consider the loop length distribution of chains.   
 
\subsection{Definition of the model}\label{sec:def} 
 
We consider a discrete model, defined on the hypercubic lattice $\mathbb Z^d$. 
Double stranded DNA segments are modeled by paths, and single stranded DNA 
segments are modeled by loops on the edges of the lattice.  Each loop is 
assumed to have two marked vertices to indicate where paths are attached. 
(Like PS, we do not {\em a priori} assume the ``DNA condition'' that the 
marked vertices divide the loop into parts of equal lengths.  There is a 
recent discussion about the effect of mismatches\mycite{GO03}.)  Any 
alternating sequence of paths and (marked) loops is called a {\it chain}.    A 
{\it PS model} consists of all chains obtained by concatenation of paths and 
loops from a given path class and a given loop class, where the initial 
segment and the final segment of a chain are both paths. (The case of an open 
end affects the behavior above the critical temperature\mycite{KMP01b} and 
will not be considered here\mycite{BCS02}). Note that, in general, such chains 
are not self-avoiding, in contrast to real DNA.  Self-avoidance may be 
violated by paths, by loops, or by the way segments are concatenated.  A chain 
is called {\it segment-avoiding} if there are no overlaps, i.e., every two 
non-neighboring segments have no vertex in common, and every two neighboring 
segments have exactly the marked vertex in common.  We call a PS model {\it 
  self-avoiding} if paths and loops are self-avoiding and if all chains of the 
model are segment-avoiding.  We will discuss below several examples of 
self-avoiding PS models, which arise from self-avoiding walks. 
 
The requirement of self-avoidance restricts the admissible path classes and 
loop classes.  A simple subclass of self-avoiding PS models are directed PS 
models:  We call a chain {\it directed} if there is a preferred 
direction such that the order of the chain segments induces the same order on 
the vertex coordinates (w.r.t. the preferred direction), for each pair of 
vertices taken from two different chain segments.  Such chains are then 
segment-avoiding.  We call a PS model {\it directed} if paths, loops and 
chains are directed.  We will discuss below several examples of directed PS 
models which arise from classes of directed walks. 
 
For a given PS model, let $z_{m,n}$ denote the number of chain configurations 
with $m$ contacts and length $n$.  The generating function is defined by 
\begin{equation}\label{form:gf} 
Z(x,w) =\sum_{m,n} z_{m,n} w^m x^n, 
\end{equation} 
where the activity $x$ is conjugate to the chain length $n$.  The Boltzmann 
factor $w=e^{-E/kT}$ takes into account the attractive interaction (achieved 
by setting the energy $E<0$) between bonds.  $T$ is the temperature, and $k$ 
is Boltzmann's constant.  Over the relevant temperature range $0< T<\infty$, 
we have $\infty>w>1$.  Note that there is no interaction between different 
segments in a chain. 
 
The generating function $Z(x,w)$ can be expressed in terms of the generating 
functions for paths $V(x)$ and loops $U(x)$. These are 
\begin{equation}\label{form:UV} 
V(x)=\sum_{n=0}^\infty b_n x^n, \qquad U(x)=\sum_{n=1}^\infty p_{2n}x^n, 
\end{equation} 
where $b_n$ is the number of paths of length $n$, and $p_{2n}$ is the number 
of loops%
\footnote{ 
In simulations of melting curves like MELTSIM\mycite{B+99,BC02}, a cooperative 
parameter $\sigma\approx 10^{-5}$ is used.  In our context, this amounts to 
replacing the loop generating function $U(x)$ by $\sigma U(x)$. 
} 
of length $2n$.  Due to the chain structure, we get a geometric series in $V(wx)U(x)$, 
\begin{equation}\label{form:PSgfun} 
Z(x,w) = \frac{V(wx)}{1-U(x)V(wx)}=\sum_{n=1}^\infty Z_n(w) x^n. 
\end{equation} 
Since we want to analyze phase transitions of the model, which can only occur 
in the infinite system, we define the free energy of the model as 
\begin{equation}\label{form:free} 
f(w)=\lim_{n\to\infty}\frac{1}{n}\log Z_n(w)=-\log x_c(w), 
\end{equation} 
where, for fixed $w$, $x_c(w)$ is the radius of convergence of $Z(x,w)$. 
Concatenation arguments and supermultiplicative inequalities can be used to 
show that the free energy exists\mycite{Ja00}.  We will alternatively 
investigate properties of the free energy in terms of the generating functions 
for paths and loops%
\footnote{ 
The reader may find it illuminating in following this very general discussion 
that now follows to refer to a specific model, discussed in Section 
\ref{sec:dir}.}%
.  Throughout the paper, we employ the following assumption. 
\begin{ass} 
Assume that the generating functions $U(x)$ and $V(x)$, defined in 
(\ref{form:UV}), have radius of convergence $x_U$ and $x_V$, respectively, 
where $0<x_U<x_V\le1$.  At the critical point $x=x_V$, assume that 
$V(x_V^-):=\lim_{x\to x_V^-}V(x)=\infty$. 
\end{ass} 
\noindent {\bf Remark.} $U(x)$ is a generating function (i.e., a series with 
non-negative coefficients), such that $U(x)$ and the derivative $U'(x)$ are 
strictly positive for $0<x<x_U$.  A corresponding statement holds for $V(x)$. 
  The assumption that $0<x_U<1$ reflects the requirement that the number of 
  configurations $p_{2n}$ grows exponentially in length.  The models commonly 
  discussed\mycite{PS66a,PS66b,F66,F84,KMP01} have $x_V=1$ and $V(0)=1$.  For 
  loops and paths of the same type, we have $x_U=x_V^2<x_V<1$, see also the 
  examples discussed below.  The assumption $V(x_V^-)=\infty$ is satisfied for 
  typical classes of paths.  Note that typically $U(x_U^-):=\lim_{x\to 
    x_U^-}U(x)<\infty$, but there are loop classes where $U(x_U^-)=\infty$ 
  such as convex polygons\mycite{BG97}.  We have $U(0)=0$ by definition. 
 
The radius of convergence $x_c(w)$ of $Z(x,w)$ is the minimum of the radius of 
convergence of $U(x)V(wx)$ and the point $x_1(w)$ where the denominator in 
(\ref{form:PSgfun}) vanishes.  Define $F(x,w):=U(x)V(wx)$.  As a function of 
argument $x$, $F(x,w)$ is continuous and monotonically increasing for 
$0<x<\min (x_U, x_V/w)$.  Note that $F(0,w)=0$.  If $w\ge x_V/x_U$, then 
$\lim_{x\to x_V/w}F(x,w)=\infty$, such that there exists a unique solution 
$x_1(w)< x_U$ with $F(x_1(w),w)=1$.  Assume that $1\le w<x_V/x_U$.  If 
$U(x_U^-)=\infty$, there exists a unique solution $x_1(w)<x_U$.  If 
$U(x_U^-)<\infty$ and $U(x_U^-)V(x_U)\ge1$, we have $F(x_U,w)\ge1$ for all 
$w\ge1$, i.e., there exists a unique solution $x_1(w)< x_U$.  If 
$U(x_U^-)<\infty$ and $U(x_U^-)V(x_U)<1$, we define $w_c>1$ by the condition 
$U(x_U^-)V(w_1\,x_U)=1$.  Then, for $w\ge w_c$, we have a unique solution 
$x_1(w)< x_U$.  If $w<w_c$, there is no such solution, and the dominant 
singularity of $Z(x,w)$ occurs at $x=x_U$.  We thus proved the following theorem. 
\begin{theorem} 
If Assumption 1 is satisfied, the PS model (\ref{form:PSgfun}) has free energy 
\begin{equation} 
f(w) = \left\{ 
\begin{array}{ll} 
-\log x_c(w) & (w_c\le w<\infty)\\ 
-\log x_U & (1<w< w_c), 
\end{array} 
\right. 
\end{equation} 
where the radius of convergence $x_c(w)$ of $Z(x,w)$ is, for $w_c\le 
w<\infty$, the unique positive solution of 
\begin{equation}\label{form:critcon} 
U(x_c(w))V(w\,x_c(w))=1. 
\end{equation} 
If $U(x_U^-):=\lim_{x\to x_U^-}U(x)=\infty$ or $U(x_U^-)V(x_U)\ge1$, we have 
$w_c=1$.  Otherwise, $w_c>1$ is implicitly given by 
$U(x_U^-)V(w_c\, x_U)=1$. \qed 
\end{theorem} 
\noindent {\bf Remark.} If $w_c>1$, this point is a critical point, see the 
following section. 
 
\subsection{Phase transitions}\label{sec:phasetr} 
 
For $w>w_c$, there is no phase transition, since the radius of convergence 
$x_c(w)>0$ is analytic for $w_c<w<\infty$, as can be inferred from 
(\ref{form:critcon}).  A necessary condition for a phase transition at a 
finite temperature $w_c>1$ is $U(x_U^-)V(x_U)<1$.  The nature of the 
transition at $w=w_c>1$ is exhibited by the fraction of shared bonds 
$\theta(w)=w\frac{df(w)}{dw}$, which is defined for $w\neq w_c$.  Note that 
$\theta(w)=0$ for $w< w_c$.  For $w>w_c$, we have 
\begin{equation} 
\theta(w)=-w\frac{x_c'(w)}{x_c(w)}=w\left(\frac{U'}{U}\frac{V}{V'}+w 
\right)^{-1}>0. 
\end{equation} 
Consider the limit $w\to w_c^+$.  We have $0<V<\infty$, since $UV=1$ and $U$ 
is finite.  Since $V(x)$ is a generating function, this implies $0<V'<\infty$. 
If $U'(x_U^-)=\infty$, it follows that $\theta(w)\to 0$, such that the phase 
transition is continuous.  If $U'(x_U^-)<\infty$, $\theta(w)\to \theta_c>0$, 
such that the phase transition is of first order.  This leads to the following 
statement. 
\begin{theorem} 
Let Assumption 1 be satisfied.  If $U(x_U^-)=\infty$ or $U(x_U^-)V(x_U)\ge1$, 
  the PS model (\ref{form:PSgfun}) has no phase transition at finite 
  temperature.  Otherwise, if $U(x_U^-)<\infty$ and $U'(x_U^-)=\infty$, a 
  continuous phase transition will occur at $w=w_c>1$ defined in Theorem 1. 
  If $U(x_U^-)<\infty$ and $U'(x_U^-)<\infty$, a first order phase transition 
  will occur at $w=w_c>1$ defined in Theorem 1. \qed 
\end{theorem} 
\noindent {\bf Remark.}  The phase transition condition $U(x_U^-)V(x_U)<1$ is 
  typically satisfied in more realistic models, where $U(x)$ is multiplied by 
  the cooperativity parameter $\sigma\approx 10^{-5}$. 
 
We conclude that the nature of the transition is determined by the singularity 
of the loop generating function $U(x)$ at $x=x_U$.  It can be more directly 
related to the asymptotic properties of loops, which are typically of the form%
\footnote{ 
$A_n \sim B_n$ for $n\to\infty$ means that $\lim_{n\to\infty} A_n/B_n=1$. 
  Similarly, $f(x) \sim g(x)$ for $x\to x_c$ means that $\lim_{x\to x_c} 
  f(x)/g(x)=1$.} 
\begin{equation}\label{form:asympt} 
p_{2n} \sim A x_U^{-n} n^{-c} \qquad (n\to\infty), 
\end{equation} 
for some constants $A>0$ and $c\in\mathbb R$.  The exponent $c$ determines the 
singularity at $x=x_U$, which is, to leading order and for $c\in\mathbb 
Q\setminus\mathbb N$, algebraic.  This specializes Theorem 2. 
\begin{prop} 
Let Assumption 1 be satisfied and assume that, for $c\in\mathbb 
R\setminus\mathbb N$, the singular part of $U(x)$ is, as $x\to x_U^-$, 
asymptotically given by 
\begin{equation}\label{form:Uc} 
U^{(sing)}(x) \sim U_0(x_U-x)^{c-1} \qquad (x\to x_U^-) 
\end{equation} 
for some constant $U_0\neq0$.  Then, the PS model (\ref{form:PSgfun}) has no 
phase transition if $0<c<1$ or if $U(x_U^-)V(x_U)\ge1$.  Otherwise, a 
continuous phase transition occurs for $1<c<2$, and a first order transition 
occurs if $c>2$, at $w=w_c>1$ defined in Theorem 1. \qed 
\end{prop} 
\noindent {\bf Remark.} If $c=1$ in (\ref{form:asympt}), we get a logarithmic 
singularity in $U(x)$, hence no phase transition.  If $c=2$ in 
(\ref{form:asympt}), we get a logarithmic singularity in the derivative 
$U'(x)$, resulting in a continuous phase transition.  See also 
Fisher\mycite{F84}. 
 
\subsection{Scaling functions}\label{sec:scalf} 
 
In the vicinity of a phase transition point, critical behavior of the form 
\begin{equation}\label{form:gsf} 
Z(x,w) \sim (x_c-x)^{-\theta} F((w-w_c)/(x_c-x)^\phi) \qquad 
(x,w)\to(x_c^-,w_c^+), 
\end{equation} 
uniformly in $w$, is expected with critical exponents $\theta$ and $\phi$. 
Here $F(s)$ is a scaling function which only depends on the combined argument 
$s=(w-w_c)/(x_c-x)^\phi$. The scaling function is extracted from the 
generating function $Z(x,w)$ by replacing $w$ using the variable $s$ of 
combined argument and expanding to leading order in $x_c-x$.  To this end, let 
us assume as in (\ref{form:Uc}) that $U(x)$ behaves as 
\begin{equation}\label{form:exp} 
U(x) = U(x_U^-) + U_0 (x_U-x)^{c-1} + {o}\left((x_U-x)^{c-1} \right) 
\qquad (x\to x_U^-), 
\end{equation} 
and $1<c\in\mathbb R\setminus\mathbb N$. In this case, the constant $U_0$ is 
negative. (If $c<1$, no phase transition occurs at positive temperature 
according to the considerations above).  We have the following result. 
\begin{theorem} 
Let Assumption 1 be satisfied.  If the leading singularity of the loop 
generating function $U(x)$ is of the form (\ref{form:exp}) with an exponent 
$c>1$ and if $U(x_U^-)V(x_U)\le1$, the PS model (\ref{form:PSgfun}) has critical 
exponents and scaling function (\ref{form:gsf}) 
\begin{equation} 
\begin{split} 
  \theta=\phi=c-1, \qquad &F(s) = \frac{1}{|U_0|- U^2(x_U^-)V'(w_c\,x_U)x_Us} 
  \qquad (1<c<2),\\ 
\theta=\phi=1, \qquad &F(s) = \frac{1}{U^2(x_U^-)V'(w_c\,x_U)[w_c-x_Us]} \qquad 
(c>2), 
\end{split} 
\end{equation} 
where $w_c\ge1$ is implicitly given by $U(x_U^-)V(w_c\, x_U)=1$.\qed 
\end{theorem} 
\noindent {\bf Remark.} If $c>2$, the critical exponents are independent of 
  $c$.  In both cases, the scaling function is proportional to 
  $F_a(s)=1/(1-as)$ with a positive constant $a$. 
 
It can be shown\mycite{Ja00} that under mild assumptions such a scaling function 
implies a certain asymptotic behavior of the function $Z_n(w)$ for large 
$n$. That is to say, 
\begin{equation} 
Z_n(w) \sim x_c^{-n-\theta}n^{\theta-1} h \left( 
\left(\frac{n}{x_c}\right)^\phi (w-w_c)\right) \qquad (n\to\infty,w\to w_c^+), 
\end{equation} 
uniformly in $w$, where $h(x)$ is the finite-size scaling function 
$h(x) = \sum_{k=0}^\infty f_k x^k/\Gamma(k\phi+\theta)$, where 
$\Gamma(z)$ denotes the Gamma function,  and the coefficients $f_k$ appear in 
the Taylor expansion of the scaling function $F(s)=\sum_{k=0}^\infty f_k s^k.$ 
This can be used to derive the critical behavior of the fraction of shared 
bonds above the phase transition\mycite{F66} for $1<c<2$, 
\begin{equation} 
\theta(w) = A (w-w_c)^\frac{2-c}{c-1} \qquad (w\to w_c^+), 
\end{equation} 
for some constant $A>0$.  In our case $F_a(s)=1/(1-as)$, we have 
$h_a(x)=e^{ax}$, if $\theta=1$.  For $0<\theta=c-1<1$ and $\theta$ rational, 
the finite size scaling function can be expressed in terms of hypergeometric 
functions.  For the case $c=3/2=\theta+1$, which will be relevant below, we 
get 
\begin{equation}\label{form:ha} 
h_a(x)=\frac{1}{\sqrt\pi}+ax 
e^{(ax)^2}\left(1+\mbox{erf}(ax)\right), 
\end{equation} 
where $\mbox{erf}(x)=2/\sqrt{\pi}\int_{0}^x e^{-t^2}\,{\rm d}t$ denotes the 
error function.   
 
\subsection{Loop statistics}\label{sec:stat} 
 
Within a given PS model, consider the set of all chains of length $n$ with $m$ 
contacts.  Denote the number of loops of length $2l$ in this set by 
$g_{m,n,l}$.  Define the generating function 
\begin{equation} 
L(x,w,y) = \sum_{m,n,l} g_{m,n,l}\,w^m x^n y^l. 
\end{equation} 
Again, due to the directedness of the model, this generating function can be 
expressed in terms of the loop generating function $U=U(x)$, $\bar U=U(xy)$ 
and the path generating function $V=V(xw)$.  An argument as in Section 
\ref{sec:def} yields 
\begin{equation} 
\begin{split} 
L(x,w,y) &= V\bar UV + (V\bar UVUV+VUV\bar UV)+\cdots\\ 
&= V\bar UV(1+2 UV + 3 (UV)^2+\cdots)\\ 
&= V\bar UV\frac{1}{(1-UV)^2} = U(yx) Z^2(x,w). 
\end{split} 
\end{equation} 
The generating function for the total number of loops is then given by setting 
$y=1$, 
\begin{equation} 
T(x,w) = L(x,w,1)=U(x) Z^2(x,w), 
\end{equation} 
and the generating function for the sum of all loop lengths is given by 
\begin{equation} 
S(x,w) = \left. y \frac{d}{dy} L(x,w,y)\right|_{y=1}=xU'(x) Z^2(x,w). 
\end{equation} 
The finite size behavior of these quantities about the phase transition can 
be computed from the scaling function results of Section \ref{sec:scalf}.  For 
the total number of loops, we get 
\begin{equation} 
\lim_{w\to w_c^+}[x^n]T(x,w) \sim A\,x_U^{-n-2\theta} n^{2\theta-1} \qquad 
(n\to\infty), 
\end{equation} 
where $A>0$ is some amplitude%
\footnote{ 
In order to simplify notation, we will denote all following amplitudes by the 
letter $A$, with the convention that their values may be different.}%
, and for the number of loops of length $2l$, we get 
\begin{equation}\label{form:loops} 
\begin{split} 
\lim_{w\to w_c^+} [x^ny^l]L(x,w,y) &\sim p_{2l} x_U^{l-n-2\theta} n^{2\theta-1} 
\qquad (n\to\infty)\\ 
&\approx A\,l^{-c} x_U^{-n-2\theta}n^{2\theta-1} \qquad (1 \ll l \ll n, \, n\to\infty). 
\end{split} 
\end{equation} 
Note that we obtained (\ref{form:loops}) under the assumption that $l$ is 
asymptotically large but that $l\ll n$.  This is due to the estimate 
$[y^l]U(yx)=p_{2l}\, x^l \sim p_{2l}x_U^l$ for $x\to x_U^-$ and $l$ finite, 
but $p_{2l}\sim A x_U^{-l}l^{-c}$ for $l\to\infty$, where we assumed 
(\ref{form:asympt}) to be valid. 
 
For the sum of the loop lengths, we distinguish between the cases $1<c<2$ and 
$c>2$, where the derivative of the loop generating function $U'(x_U^-)$ is 
finite or infinite at the critical point.  We find 
\begin{equation} 
\lim_{w\to w_c^+} [x^n] S(x,w) \sim\left\{ 
\begin{array}{ll} 
A\,x_U^{-n-c} n^{c-1}& (1<c<2)\\ 
A\,x_U^{-n-2} n & (c>2) 
\end{array} 
\right. \qquad (n\to\infty). 
\end{equation} 
Let us compute the probability of a loop of length $2l$ within chains of 
length $n$.  We get, in the limit $w\to w_c^+$, for the loop length 
distribution 
\begin{equation}\label{form:PSlld} 
P(l,n) = \lim_{w\to w_c^+} \frac{[x^ny^l]L(x,w,y)}{[x^n]T(x,w)} \approx A l^{-c} 
\qquad (1 \ll l \ll n, \, n\to\infty). 
\end{equation} 
Furthermore, for the mean loop length $\langle\, l \,\rangle_n$ we get, in the 
limit $w\to w_c^+$, 
\begin{equation}\label{form:PSmean} 
\langle\, l \,\rangle_n = \sum_{l=0}^n l\,  P(l,n) = \lim_{w\to w_c^+} 
\frac{[x^n]S(x,w)}{[x^n]T(x,w)}\sim\left\{ 
\begin{array}{ll} 
A\,n^{2-c}& (1<c<2)\\ 
A\,n^0 & (c>2) 
\end{array} 
\right. \qquad (n\to\infty). 
\end{equation} 
The behavior of the mean loop length reflects the nature of the phase 
transition:  If $1<c<2$, the mean loop length diverges, such that the 
transition is continuous.  If $c>2$, the mean loop length  stays finite, 
indicating a first order transition. 
 
\section{Two prominent loop classes} 
 
Since the critical behavior of PS models is essentially determined by the
properties of loops, PS\mycite{PS66a,PS66b}, and later Fisher\mycite{F66},
were led to consider various loop classes (together with straight paths for
the double stranded segments).  Whereas PS analyzed loop classes derived from
random  walks, Fisher considered loop classes derived from self-avoiding walks.   
 
\subsection{Loops and walks}\label{sec:lw} 
 
An oriented, rooted loop of length $n$ is a walk of length $n-1$, whose
starting point and end point are lattice nearest neighbors.  As
usual\myccite{Sec.~3.2}{MS93}, we identify such loops if they have the same
shape, i.e., if they are equal up to a translation, possibly followed by a
change of orientation.  These objects we call unrooted, unoriented loops, or
simply loops.  Each loop of length $n$ has at most $2n$ corresponding walks.
If the walks are self-avoiding, each loop has exactly $2n$ corresponding walks.
The number of loops of length $n$ is denoted by $p_n$.  For example, for
self-avoiding loops on $\mathbb Z^2$ we have $p_4=1$ and $p_6=2$.
For a given class of walks, the above description defines the corresponding 
(unmarked) loop class.  Within a chain structure, two paths are attached to 
each loop.  Different choices for attachment positions increase the number of 
(marked) loop configurations to ${\widetilde p}_n\ge p_n$.  If we assume the 
DNA condition that the two paths attached to a loop bisect it into pieces of 
equal length, then the number of possible attachments of two paths to a loop 
of length $n$ is less than or equal to $2n$.  (We distinguish the two strands of a 
marked loop).  PS and Fisher consider classes of oriented rooted loops.
Self-avoiding oriented rooted loops can be interpreted as loops with $2n$ possible 
attachment positions of paths to a loop of length $n$. A similar interpretation
for oriented rooted random loops is not obvious.  We stress that these loop classes 
result in chains which are not segment-avoiding, as paths will intersect the 
loops.  Both models cannot therefore represent real (self-avoiding) DNA. 
 
\subsection{Oriented rooted random loops} 
 
The first simple example of loops, which are discussed by PS\mycite{PS66b}, 
are oriented rooted random loops derived from random walks.  Random walks on 
$\mathbb Z^d$ have the generating function $V_d(x)=1/(1-2dx)$.  The asymptotic 
behavior of the number of oriented rooted random loops of length $2n$ is 
given\myccite{App.~A}{MS93} by 
\begin{equation} 
{\widetilde p}_{2n}\sim A (2d)^{2n} (2n)^{-d/2} \qquad (n\to\infty). 
\end{equation} 
This implies no phase transition in $d=2$ and a continuous phase transition in 
$d=3$, since the phase transition condition $U(x_U^-)V(x_U)<1$ is satisfied in 
$d=3$:  We have 
\begin{equation} 
{\widetilde p}_{2n}=\sum_{k+l+m=n} \frac{(2n)!}{(k!)^2\,(l!)^2\,(m!)^2}= 
\frac{((2n)!)^2}{(n!)^4}\, {}_3F_2(-n,-n,-n;1,-n+1/2;1/4), 
\end{equation} 
where ${}_3F_2(a_1,a_2,a_3;b_1,b_2;z)$ is a hypergeometric function.  With 
$x_U=1/36$, we extracted the amplitude $U(x_U^-)$ numerically\mycite{G89} 
using the values ${\widetilde p}_{2n}$ where $n\le30$, getting 
$U(x_U^-)=0.51638461326(7)$.  The result follows for straight paths since 
$1/(1-x_U)=36/35$ and also for random walks since $V_3(x_U)=6/5$. 
 
\subsection{Oriented rooted self-avoiding loops}\label{sec:rSAP} 
 
The PS results led to the question\mycite{F66} whether accounting for excluded 
volume effects within a loop increases the loop class exponent $c$, which 
might change the order of the phase transition.  This led to considering 
{\em self-avoiding loops}, which are loops derived from self-avoiding walks. 
(A self-avoiding walk\myccite{Sec.~1.1}{MS93} is a random walk which never visits a 
vertex twice.)  By definition, the self-avoiding loop class fully accounts 
for excluded volume interactions within a loop.  In $d = 3$, self-avoiding
loops of length $n\ge24$ may be knotted.  Self-avoiding walks and loops are well studied
objects\mycite{MS93,Ja00}. Fisher considered oriented rooted loops
${\widetilde p}_{2n}=4np_{2n}$.  Their loop class exponent $c$, ${\widetilde
  p}_{2n}\sim B \mu_d^n n^{-c}$, is related to the mean square displacement
exponent $\nu$ of self-avoiding walks by the hyperscaling
relation\myccite{Sec.~2.1}{MS93} $c=d\nu$, where $\nu=1/2$ for $d\ge4$,
$\nu=0.5877(6)$ for $d=3$ and $\nu=3/4$ for $d=2$.  At present, there is no
proof for the values of $\nu$ in dimensions $d\le4$. Explicitly, we have 
\begin{equation} 
c= 
\left\{ 
\begin{array}{ll} 
d/2&d\ge4\\ 
1.7631(18)&d=3\\ 
3/2&d=2\\ 
\end{array} 
\right. 
\end{equation} 
For unknotted self-avoiding loops, which is the preferrable model from a
biological point of view, it has been proved\mycite{SW88} that the exponential
growth constant is strictly less than that of all self-avoiding loops, while
the exponent (if it exists) is expected to coincide with that of all
self-avoiding loops. 

Fisher concluded that the above values of the loop class exponent $c$ imply a
continuous phase transition in $d=2$ and $d=3$.  We have to check the phase
transition condition:  In $d=2$, a numerical analysis of the oriented rooted
SAP series data\mycite{J03} gives $x_U=0.1436806285(8)$ and
$U(x_U^-)=0.6523866(2)$.  The value of the generating function for straight
paths is $1/(1-x_U)<1.678$.  In $d=3$, a corresponding analysis with unknotted
oriented rooted SAP data and SAW data\myccite{App.~C}{MS93} gives $x_U=0.045578(3)$
and $U(x_U^-)=0.10(1)$.  The path generating function value is bounded by that
of random walks $1/(1-6x_U)<1.38$.  Thus, the phase transition condition is
satisfied in both $d=2$ and in $d=3$ for straight paths.  Note that in $d=2$,
self-avoiding walks as paths will result in no phase transition, since
$V(x_U)>2$ for such paths, as follows from SAW data analysis.
 
\section{Self-Avoiding Poland-Scheraga models}\label{sec:sPS} 
 
As discussed above, the previous PS models are not self-avoiding.  Given a 
particular class of walks, for example, SAW, a PS model with segment-avoiding 
chains may be defined as follows.  As paths, take only those walks with 
extremal first and last vertex: If $v(0),v(1),\ldots, v(n)$ are the vertices 
of an $n$-step walk $v$, this walk is taken as a path iff 
$v_x(0)<v_x(i)<v_x(n)$ for all $1<i<n$.  Such walks are 
bridges\myccite{Sec.~1.2}{MS93}, whose last step is in $x$-direction.  For 
(unmarked) loops, take loops derived from the walks, as explained in Section 
\ref{sec:lw}. In our example these will be SAW loops.  Marking of the loops may 
be achieved in different ways.  To this end, consider for a given loop the 
sets $A_l$ ($A_r$) of vertices of smallest (largest) $x$-coordinate. We 
distinguish four different types of marking:  complete marking (with DNA 
constraint), where we mark a loop at all vertex pairs from $A_l$ and $A_r$ 
(whenever the DNA condition is satisfied), and unique marking (with DNA 
constraint), where we only mark a loop at a single vertex pair, for example 
the bottom vertex and the top vertex in a lexicographic ordering (if they 
satisfy the DNA condition). 
 
If the walk class is self-avoiding walks, this will result in self-avoiding PS 
models.  Unique marking would then imply ${\widetilde p}_n=2p_n$ (we 
distinguish the two strands of a marked loop) and hence increase the previous 
exponents by one.  Hence, such a PS model displays a first order transition 
with $c=5/2$ in $d=2$ and with $c=2.7631(18)$ in $d=3$.  (The phase transition 
condition is satisfied, as follows from the values $U(x_U^-)$ given in Section 
\ref{sec:rSAP}, and from the estimate $V(x_U)<1/(1-dx_U)-1$ for bridges with 
last step in $x$-direction.)  Unique marking with DNA constraint results in 
${\widetilde p}_n\le 2p_n$.  If we assume that the exponential growth constant 
for marked self-avoiding loops is still given by $\mu_d$ (which we expect to 
be true, compare Section \ref{sec:part}), this implies a critical exponent $c$ 
greater or equal to the model with unique marking, i.e., a first order phase 
transition in $d=2$ and $d=3$.  For complete marking, we have 
$2p_n\le{\widetilde p}_n\le 2n p_n$, which rules out a decrease of $c$ by more 
than one.  We however expect that the number of possible markings is of order 
$1$ as $n\to\infty$, such that $c$ remains unchanged, and the model again 
displays a first order phase transition.  Similar considerations apply for the 
case of complete marking with DNA constraint. 
 
\section{Directed Poland-Scheraga models}\label{sec:exact} 
 
We present two classes of directed PS models which are exactly solvable and, 
by definition, take into account excluded volume interactions between all 
parts of the structure.  The first class is derived from fully directed walks 
and is solvable in arbitrary dimension\mycite{GP93}.  We will give explicit 
expressions for the generating function in $d=2$ and in $d=3$.  This 
extends\mycite{L91, KS92, MBM01, MBM02}, where the critical behavior has been 
derived.  The model displays a first order phase transition only for $d>5$. 
The second class, derived from partially directed walks, is considered in 
$d=2$ only.  Different variants of the second model display both a first order 
phase transition and a continuous phase transition in $d=2$. 
 
\subsection{Fully directed walks and loops}\label{sec:dir} 
 
This directed PS model consists of fully directed walks for the paths in the 
chain.  These only take steps in positive directions.  The corresponding loops 
are staircase polygons, which consist of two fully directed walks, which do 
not intersect or touch, but have a common starting point and end point.  Paths 
are attached to these points.  We distinguish the two strands of a loop.  This 
model satisfies all the assumptions discussed in Section \ref{sec:def}, and 
also satisfies the DNA condition that the two segments of the loops are the 
same length. 
 
In $d=2$, the generating functions for paths and marked loops are\mycite{Ja00} 
\begin{equation} 
V(x) =\frac{1}{1-2x}, \qquad U(x)=1-2x-\sqrt{1-4x}. 
\end{equation} 
$U(x)$ is twice the generating function of staircase polygons.  The loop class
exponent is $c=3/2$.  We have  $x_V=1/2$, $x_U=1/4$ and $w_c=1$, i.e., a phase
transition at $T=\infty$.  (If the empty path would not be allowed, a phase
transition occurred at a finite  temperature.)  The free energy $f(w)$ is given by 
\begin{equation} 
f(w) =  \log\left( \frac{2(w-1)^2}{\sqrt{1+(w-1)^2}\,-1}\right)  \qquad (1\le 
w<\infty). 
\end{equation} 
The fraction of shared bonds follows as 
\begin{equation}\label{form:2dtheta} 
\theta(w)=  
\frac{2w}{w-1}\frac{\sqrt{1+(w-1)^2}-1-(w-1)^2/2} 
{1+(w-1)^2-\sqrt{1+(w-1)^2}} 
\qquad (1\le w<\infty), 
\end{equation} 
which approaches zero linearly in $w-1$.  The asymptotic behavior of 
$Z_n(w)$ about $w=1$ is given by 
\begin{equation} 
Z_n(w) \sim \frac{4^{n-1}}{n^{1/2}} h_{\sqrt{n}/8}(w-1) \qquad 
(n\to\infty, \, w\to 1^+), 
\end{equation} 
uniformly in $w$, where $h_a(x)$ is given by (\ref{form:ha}). 
 
The PS model of fully directed walks and loops is exactly solvable in 
arbitrary dimension\mycite{GP93}.  Consider fully directed walks on $\mathbb 
Z^d$.  If there are $k_i$ steps in direction $i$, the number of distinct 
walks, starting from the origin, is given by the multinomial coefficient 
$\binom{k_1+k_2+\ldots+k_d}{k_1,k_2,\ldots,k_d}$.  This results in 
$V_d(x)=1/(1-dx)$.  The generating function of a pair of such walks is given by 
\begin{equation} 
{\widetilde Z}_d(x)=\sum_{k_1,k_2,\ldots,k_d=0}^\infty 
\binom{k_1+k_2+\ldots+k_d}{k_1,k_2,\ldots,k_d}^2 x^{k_1+\ldots+k_d}. 
\end{equation} 
The generating function ${\widetilde Z}_d(x)$ can be interpreted as a chain, 
where each link consists of a staircase polygon or a double bond.  Thus, it is 
related to the generating function $U_d(x)$, where staircase polygons are 
counted twice, by 
\begin{equation} 
{\widetilde Z}_d(x) = \frac{1}{1-(dx+U_d(x))}. 
\end{equation} 
The functions ${\widetilde Z}_d(x)$ satisfy Fuchsian differential 
equations of order $d-1$, from which the singular behavior of ${\widetilde 
  Z}_d(x)$, and hence of $U_d(x)$ can be derived\mycite{GP93}.  In dimensions 
$2<d<5$, $U_d(x)$ can be expressed in terms of Heun functions.  We get in 
$d=3$ 
\begin{equation} 
{\widetilde Z}^2_3(x)=\left(\frac{2}{\pi}\right)^2 (1-9x)^{-1}(1-x)^{-1}K(k_+)K(k_-), 
\end{equation} 
where $K(k)$ is the complete elliptic integral of the first kind, and 
\begin{equation} 
\begin{split} 
k_\pm^2&=\frac{1}{2}\pm 
\frac{x_1}{4}(4-x_1)^\frac{1}{2}-\frac{1}{4}(2-x_1)(1-x_1)^\frac{1}{2} \\ 
 x_1&=-\frac{16x}{9 x^2-10x+1} 
\end{split} 
\end{equation} 
We have $x_U=1/9$.  In arbitrary dimension $d\ge3$, it has been 
shown\mycite{GP93} that the models have a critical point $x_U=1/d^2$ with 
exponent $c=(d-1)/2$, with logarithmic corrections in $d=3$. 
Noting that $U_d(x_U^-)=1-1/d-1/{\widetilde Z}_d(x_U^-)$ and 
$V_d(x_U)=d/(d-1)$, we conclude that the phase transition condition 
$U_d(x_U^-)V_d(x_U)<1$ is satisfied iff ${\widetilde Z}_d(x_U^-)<\infty$. 
This is the case\mycite{GP93} in $d\ge4$.  For the corresponding PS model, the 
phase transition is thus first order for $d\ge6$, and we have a continuous phase 
transition in dimensions $2\le d\le5$, being at finite temperature in $d\ge4$ only. 
 
\subsection{Partially directed walks and loops}\label{sec:part} 
 
We consider a directed PS model, where paths are partially directed walks on 
the square lattice $\mathbb Z^2$.  These walks are self-avoiding and not 
allowed to take steps in the negative $x$ direction.  The number of paths of 
length $n$ can be obtained\mycite{Ja00} by considering the ways a walk 
of length $n$ can be obtained from a walk of length $n-1$.  This yields the 
path generating function 
\begin{equation} 
V(x) =\frac{1+x}{1-2x-x^2}, 
\end{equation} 
whose dominant singularity is a simple pole at $x_V=1-\sqrt{2}$.  The 
associated loop class is column-convex polygons, which has been analyzed 
in\mycite{BOP94}, see Fig.~\ref{fig:cc} for an example. 
%
\Bild{4.5}{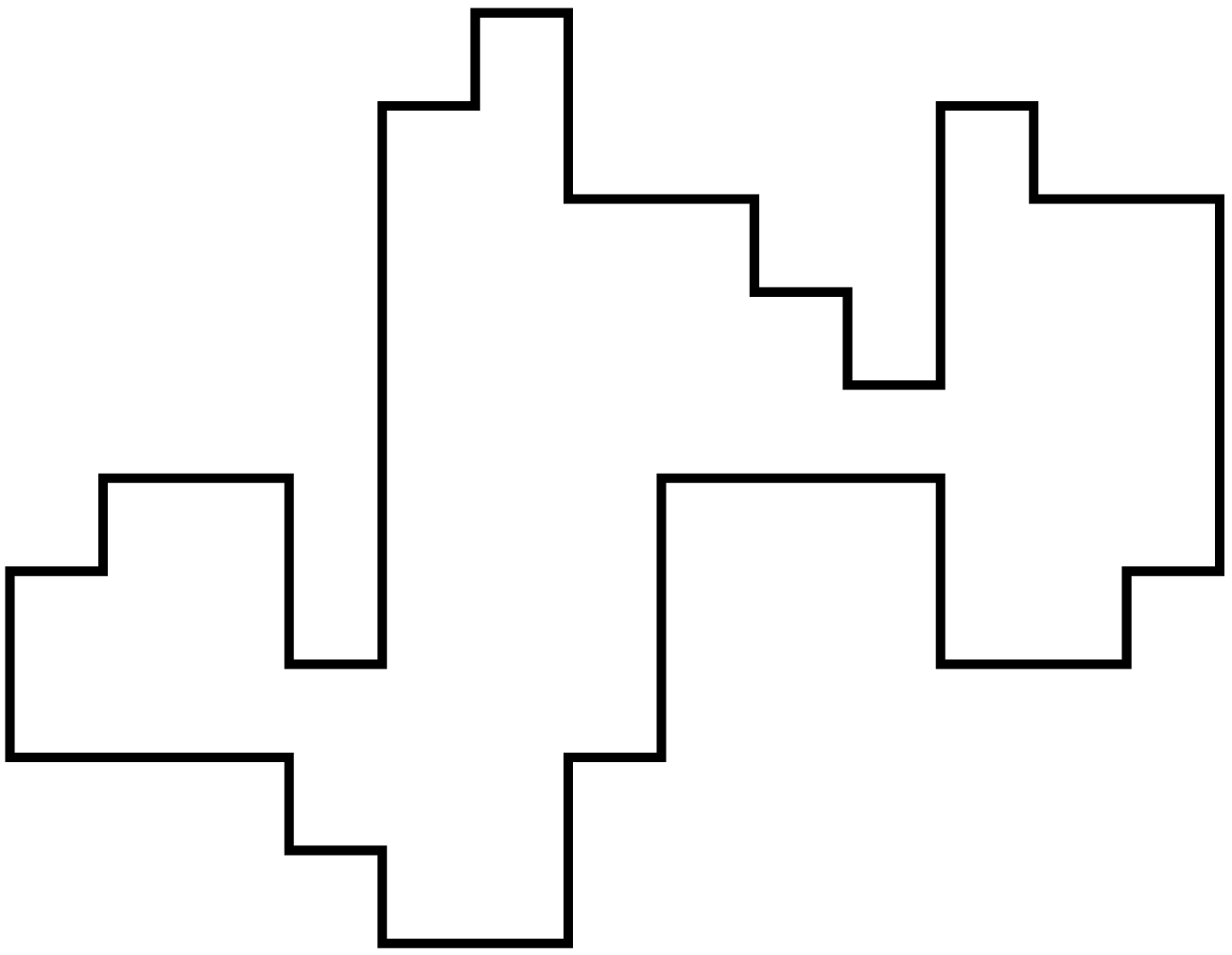}{A column-convex polygon}{fig:cc} 
We assume unique marking as explained in Section \ref{sec:sPS}.  The 
(unmarked) loop generating function $U(x)$ satisfies the algebraic 
equation of order four\myccite{Eqn.~5.2}{BOP94} 
\begin{equation}\label{form:cc} 
\begin{split} 
0=&-(x-1)^4x^2+(x^3-7x^2+3x-1)(x-1)^3\,U(x)\\ 
&+2(2x^3-11x^2+10x-4)(x-1)^2\,U(x)^2\\ 
&+(5x^3-35x^2+47x-21)(x-1)\,U(x)^3\\ 
&+(2x^3-23x^2+38x-18)\,U(x)^4. 
\end{split} 
\end{equation} 
The dominant singularity at $x_U=x_V^2=3-2\sqrt{2}$ is a square-root 
singularity with an exponent $c=3/2$.  The critical amplitude is given by 
$U(x_U^-)=0.086983(1)$.  Thus $2U(x_U^-)V(x_U)<1$. This implies a continuous 
phase transition for the PS model at $w_c=1.922817(1)$. 
 
However, the loop class defined above does not obey the DNA constraint, which 
might increase the exponent $c$.  Using standard techniques\mycite{BOP94}, it
is possible to derive a functional equation for the generating function of
column-convex polygons counted by perimeter and, in addition, by their upper
and lower walk length.  The corresponding expression is, however, much more
difficult to handle than (\ref{form:cc}), and there seems to be no practical
way to extract the loop class exponent for the model with DNA constraint.  On
the other hand, it is possible to bound the loop class exponent from above by
considering a refined (artificial) model.  As we will argue below, this
refined model has a loop class exponent $c=5/2$.  For the former model, we
thus expect a value of $3/2\le c\le5/2$, with a value of $c=2$ seeming
plausible.  This PS model would then be at the border between a first order
and a continuous phase transition.
 
For the refined model, let us further assume that the upper and lower walk in 
a loop are not allowed to touch the $x$-axis.  The loop class of the refined model can be 
described in terms of bar-graph polygons\mycite{PB95}.  These are 
column-convex polygons, one of whose walks is a straight line.  Loops are 
composed of two bar-graph polygons, one of them in the half-plane $y\ge0$, the 
other in the half-plane $y\le0$.  Both polygons have the same horizontal 
coordinates and are constrained to have equal walk lengths.  Paths of the 
model are partially directed walks, which are attached to the two loop 
vertices with vertical coordinate zero. 
 
Let $g_{m,n}$ denote the number of bar-graph polygons with $2m$ horizontal and 
$2n$ vertical steps.  The anisotropic perimeter generating function 
$G(x,y)=\sum g_{m,n}x^my^n$ satisfies an algebraic equation of degree 
two\mycite{PB95}, 
\begin{equation} 
G(x,y)=xy+(y+x(1+y))G(x,y)+xG^2(x,y). 
\end{equation} 
A generalization of the Lagrange inversion formula\mycite{FS01} can be used to 
obtain a closed formula for its series coefficients.  It is 
\begin{equation} 
[x^my^n]G(x,y)=g_{m,n}=\sum_{k=0}^n 
\frac{1}{m+k}\binom{m+k}{m}\binom{m}{n-1-k}\binom{m}{n-k}. 
\end{equation} 
The number $p_{2n}$ of all combinations of bar-graph polygons subject to equal 
walk lengths is then 
\begin{equation} 
p_{2n}=\sum_{k=1}^{(n-1)/2} g_{n-2k,k}^2. 
\end{equation} 
The first few numbers $p_{2n}$ are $1,1,2,10,38,126,483,126,483,\ldots$ for 
$n=3,4,5,\ldots$.  We used standard methods of numerical series 
analysis\mycite{G89} to estimate the critical point and critical exponent of 
$U(x)$.  An analysis with first order differential approximants, using the 
coefficients $p_{2n}$ for $70\le n\le 80$, yields the estimates 
\begin{equation} 
x_U = 0.17157287(1), \qquad c=2.5000(1) 
\end{equation} 
This is, to numerical accuracy, indistinguishable from 
$x_U=3-2\sqrt{2}=0.171572875253\ldots$, as expected, and $c=5/2$.  Note that 
the phase transition condition $2U(x_U^-)V(x_U)<1$ is satisfied, since the 
loop generating function is bounded by that of column-convex polygons, which 
is known to satisfy the above inequality. 
 
\section{Models extending the Poland-Scheraga class} 
 
The PS model with oriented rooted self-avoiding loops\mycite{F66}, reviewed in
Section \ref{sec:rSAP}, displays a continuous phase transition in $d=2$ and
$d=3$.  This led to the question whether introducing excluded volume effects
between different segments of the chain can change the nature of the
transition\mycite{CCG00, KMP00, KMP01, KMP02,COS02, BCS02, BCKMOS02}.  In
Section \ref{sec:sPS}, we introduced self-avoiding PS models in order to
answer that question in the affirmative, staying within the PS class.
 
Another approach, which extends the PS class, consisted in simulations of
self-avoiding walk pairs\mycite{CCG00, COS02, BCS02, BCKMOS02}.  We emphasize
that these models are not PS models as defined in this paper, and that loop
class exponents as defined in (\ref{form:asympt}) bear no meaning for these
models.  The appropriate generalization is the exponent $c'$ of the loop
length distribution $P(l,n)$ of the chain, which is assumed to behave in the
limit $w\to w_c^+$ like\mycite{KMP00, KMP01, KMP02, COS02, BCS02, BCOS02}
\begin{equation}\label{form:pdf} 
P(l,n) \approx l^{-c'} g(l/n) \qquad (1\ll l\le n,\, n\to\infty), 
\end{equation} 
where $g(x)$ is a scaling function, assumed to be constant in some 
analyses\mycite{KMP00, KMP01, KMP02, BCOS02}.  Then, the same conclusions as in 
Section \ref{sec:phasetr} about the nature of the phase transition determined 
by the loop length distribution exponent $c'$ hold.  This follows from the 
behavior of the mean loop length $\langle\, l \,\rangle_n$ in the limit $w\to 
w_c^+$, which is obtained from (\ref{form:pdf}) by integration, 
\begin{equation}\label{form:mean} 
\langle\, l \,\rangle_n = \sum_{l=0}^n l\,  P(l,n) \approx A\,n^{2-c'} 
\qquad (n\to\infty). 
\end{equation} 
If $1<c'<2$, the mean loop length diverges, indicating a continuous transition. 
If $c'>2$, the mean loop length stays finite, indicating a first order 
transition\mycite{COS02,BCOS02}.  Note that assumption (\ref{form:pdf}) is not 
fully consistent with PS models: The behavior (\ref{form:PSlld}) is 
consistent with (\ref{form:pdf}), which justifies the use of the same name for 
the two exponents.  We found (\ref{form:mean}) to be satisfied for PS models 
if $1<c<2$ in Section \ref{sec:stat}.  If $c>2$, however, the mean loop length 
exponent is independent of $c$ for PS models---see (\ref{form:PSmean}).  This 
implies that a scaling form like (\ref{form:pdf}) cannot hold for PS models 
with loop class exponent $c>2$. 
 
The simulation results are $c'=2.44(6)$ in $d=2$\mycite{BCS02} and $c'=2.14(4)$ in 
$d=3$\mycite{BCKMOS02}, indicating a first order phase transition in $d=2$ and 
in $d=3$.  This led to the conclusion that introducing excluded volume 
effects between different segments to the PS model of rooted self-avoiding 
loops drives the transition from continuous to first order\mycite{KMP00, KMP01, 
  KMP02}.  The perturbative analytic arguments using the theory of 
polymer networks\mycite{KMP00, KMP01, KMP02} yield a very good 
approximation to the value of the loop length distribution exponent found in 
simulations.  This suggests that the main mechanism responsible for the change 
of critical behavior in the PS model of oriented rooted self-avoiding loops is
due to ``local'' excluded volume effects arising from forbidden attachments of
paths to loops.

\section{Conclusion} 
 
We discussed PS models with emphasis on the order of their phase transitions. 
We re-analyzed the old model of oriented rooted self-avoiding loops\mycite{F66} 
and found that the resulting PS model is not self-avoiding. Hence the 
conclusions about the order of the phase transition, which are most relevant 
to the recent discussions\mycite{CCG00, KMP00, KMP01, KMP02, COS02, BCOS02, 
  BCS02, BCKMOS02, HM01, KMP01b}, rely on a model where most chain 
configurations are not self-avoiding and thus cannot represent real DNA. Our 
self-avoiding PS model (unique marking, with self-avoiding bridges and 
(unrooted) self-avoiding loops as defined in Section \ref{sec:sPS}) yields a 
first order phase transition in both $d=2$ and $d=3$.  It is expected that the 
other variants considered in Section \ref{sec:sPS}  also yield a first order 
phase transition, but a detailed (numerical) analysis is needed to answer this 
question, as in the discussion in Section \ref{sec:sPS}. 
 
For the presumably more realistic model of pairs of interacting 
self-avoiding walks\mycite{CCG00, KMP00, KMP01, KMP02, COS02, BCS02, 
  BCKMOS02}, our results suggest an interpretation of excluded volume effects, 
which complements the common one\mycite{KMP00, KMP01, KMP02}.  The 
self-avoiding PS models of Section \ref{sec:sPS} correctly account for 
excluded volume effects within a loop, but overestimate excluded volume 
effects between different segments of a chain, due to their directed chain 
structure.  Since this leads to a first order phase transition in $d=2$ and 
$d=3$, we conclude that the relaxation of excluded volume effects between 
different segments of the chain does not change the nature of the transition. 
 
We also discussed several directed examples.  One of these models displays a 
first order phase transition in $d=2$.  Despite being exactly solvable, these 
models seem to be of limited relevance to the biological problem due to 
their directed structure and other limitations. 
 
The key question, as to which effects are responsible for the observed 
behavior in real DNA, is only partially answered by these results.  With 
respect to directed models, which also account for different base pair 
sequences, it seems surprising however that simulations of melting 
curves\mycite{WB85,B+99} agree so well with experimental curves.  In these 
simulations, a heuristic critical exponent accounting for the statistical 
weight of internal loops has to be inserted.  It has recently been 
suggested\mycite{BC02} that a value more realistic than the commonly used 
Fisher loop class exponent\mycite{F66} $c\approx1.763$ might be the loop 
length distribution exponent of the polymer network 
approximation\mycite{KMP00} $c'\approx2.115$.  This suggestion implicitly 
assumes that the loop length distribution exponent of a non-directed 
model can be interpreted as an ``effective'' loop class exponent within a 
directed model.  Although appearing plausible, it seems difficult to give this 
assumption a quantitative meaning.  On the other hand, the MELTSIM simulation 
approach\mycite{B+99} is robust against a change of parameters.  Indeed, 
melting curves can be simulated with satisfactory coincidence for both 
exponents, if the cooperativity parameter for the loops is adjusted 
accordingly\mycite{BC02}. It would certainly be illuminating to obtain such 
simulated melting curves from a self-avoiding walk pair modeling\mycite{CCG00, 
  COS02, BCS02, BCKMOS02}.  We also mention the recent debate about the 
relevance of the self-avoiding walk pair model to real DNA\mycite{HM01,KMP01b}. 
 
The stiffness of the double stranded segments does not seem to qualitatively 
alter the critical properties of chains\mycite{COS02}, as can be seen by 
comparing the PS model result of Section 2.1, showing that critical properties 
are largely independent of the path generating function.  The corresponding 
question of introducing energy costs for bending and torsion to single 
stranded segments has, to our knowledge, not been discussed.  Even within the 
PS class, it is not clear how this affects the loop class exponents, since the 
changes just discussed might lead to loop classes with different exponential 
growth constants, but there is no argument as to how the exponents might change. 
 
In conclusion, the question of the mechanisms applying in real DNA which are 
responsible for the denaturation process and which explain multistep behavior 
as observed in melting curves, are still far from being satisfactorily 
answered in our opinion. 
 
\section*{Acknowledgments} 
 
We would like to acknowledge useful discussions with Stu Whittington.  CR 
would like to acknowledge financial support by the German Research Council 
(DFG) and like to thank the Erwin Schr\"odinger International Institute for 
Mathematical Physics for support during a stay in December 2002, where part of 
this work was done.  AJG would like to acknowledge financial support from the 
Australian Research Council, and helpful discussions with Marianne Frommer. 
We acknowledge clarifying comments on an earlier version of the manuscript by 
Marco Baiesi, Enrico Carlon, Andreas Hanke and Ralf Metzler.


\begin{thebibliography}{99} 
 
\bibitem{BCKMOS02} 
M.~Baiesi, E.~Carlon, Y.~Kafri, D.~Mukamel, E.~Orlandini and 
A.~L.~Stella, 
Inter-strand distance distribution of DNA near melting, 
{\it Phys.~Rev.~E \bf 67} (2002), 21911--21917; 
{\em cond-mat/0211236}. 
 
\bibitem{BCOS02} 
M.~Baiesi, E.~Carlon, E.~Orlandini and A.~L.~Stella, 
A simple model of DNA denaturation and mutually avoiding walk statistics, 
{\em Eur.~Phys.~J.~B \bf 29} (2002), 129--134; 
{\em cond-mat/0207122}. 
 
\bibitem{BCS02} 
M.~Baiesi, E.~Carlon and A.~L.~Stella, 
Scaling in DNA unzipping models: denaturated loops and end-segments as 
branches of a block copolymer network, 
{\it Phys.~Rev.~E \bf 66} (2002), 21804--21812; {\it cond-mat/0205125}. 
 
\bibitem{B+99} 
R.~D.~Blake et al., 
Statistical mechanical simulation of polymeric DNA melting with 
MELTSIM, 
{\it Bioinformatics \bf 15} (1999), 370--375. 
 
\bibitem{BC02} 
R.~Blossey and C.~Carlon, 
Reparametrizing loop entropy weights: effect on DNA melting curves, 
{\it preprint} (2002); {\it cond-mat/0212457}.

\bibitem{BG97}
M.~Bousquet-M\'elou and A.~J.~Guttmann,
Enumeration of three-dimensional convex polygons,
{\it Ann.~Comb. \bf 1} (1997), 27--53.
 
\bibitem{BOP94} 
R.~Brak, A.~L.~Owczarek and T.~Prellberg, 
Exact scaling behavior of partially convex vesicles, 
{\em J.~Stat.~Phys. \bf 76} (1994), 1101--1128. 
 
\bibitem{COS02} 
E.~Carlon, E.~Orlandini and A.~L.~Stella, 
The roles of stiffness and excluded volume in DNA denaturation, 
{\it Phys.~Rev.~Lett. \bf 88} (2002), 198101--198104; {\it 
cond-mat/0108308}. 
 
\bibitem{CCG00} 
M.~S.~Causo, B.~Coluzzi and P.~Grassberger, 
Simple model for the DNA denaturation transition, 
{\em Phys.~Rev.~E \bf 62} (2000), 3958--3973. 
 
\bibitem{F66} 
M.~E.~Fisher, 
Effect of excluded volume on phase transitions in biopolymers, 
{\em J.~Chem.~Phys. \bf 45} (1966), 1469--1473. 
 
\bibitem{F84} 
M.~E.~Fisher, 
Walks, walls, wetting, and melting, 
{\it J.~Stat.~Phys. \bf 34} (1984), 667--729. 
 
\bibitem{FS01} 
P.~Flajolet and R.~Sedgewick, 
Analytic combinatorics: Functional equations, rational and algebraic 
functions, {\em INRIA preprint 4103} (2001). 
 
\bibitem{GO03} 
T.~Garel and H.~Orland, 
On the role of mismatches in DNA denaturation, 
{\it Saclay preprint T03/045} (2003); {\it cond-mat/0304080}. 
 
\bibitem{G89} 
A.~J.~Guttmann, Asymptotic analysis of power-series expansions, in: 
{\it Phase Transitions and Critical Phenomena} vol.~13 eds. C.~Domb 
and J.~L.~Lebowitz, Academic, New York (1989), pp. 1--234. 
 
\bibitem{GP93} 
A.~J.~Guttmann and T.~Prellberg, 
Staircase polygons, elliptic integrals, Heun functions, and lattice 
Green functions, 
{\it Phys.~Rev.~E \bf 47} (1993), R2233--R2236. 
 
\bibitem{HM01} 
A.~Hanke and R.~Metzler, 
Comment on ``Why is the DNA denaturation transition first order?'' 
{\it Phys.~Rev.~Lett. \bf 90} (2003) 159801; {\em cond-mat/0110164}. 
 
\bibitem{HM02} 
A.~Hanke and R.~Metzler, 
Entropy loss in long-distance DNA looping {\it preprint} (2002); 
{\em cond-mat/0211468}. 
 
\bibitem{Ja00}  
E.~J.~Janse van Rensburg, 
{\it The Statistical Mechanics of Interacting walks, Polygons, Animals 
and Vesicles}, Oxford University Press, New York (2000). 
 
\bibitem{J03}  
I.~Jensen, 
A parallel algorithm for the enumeration of self-avoiding polygons on the 
square lattice, 
{\it J.~Phys.~A:~Math.~Gen. \bf 36} (2003), 5731--5745;  
{\em cond-mat/0301468}. 
 
\bibitem{KMP00} 
Y.~Kafri, D.~Mukamel and L.~Peliti, 
Why is the DNA denaturation transition first order?, 
{\em Phys.~Rev.~Lett. \bf 85} (2000), 4988--4992; 
{\em cond-mat/0007141}. 
 
\bibitem{KMP01} 
Y.~Kafri, D.~Mukamel and L.~Peliti, 
Melting and unzipping of DNA, 
{\em Eur.~Phys.~J. B \bf 27} (2001), 135--146; 
{\em cond-mat/0108323}. 
 
\bibitem{KMP01b} 
Y.~Kafri, D.~Mukamel and L.~Peliti, 
Reply to comment by Hanke and Metzler 
{\it preprint} (2001); {\em cond-mat/0112179}, {\em cond-mat/0302589}. 
 
\bibitem{KMP02} 
Y.~Kafri, D.~Mukamel and L.~Peliti, 
Denaturation and unzipping of DNA: statistical mechanics of 
interacting loops, 
{\it Physica A \bf 306} (2002), 39--50. 
 
\bibitem{KS92} 
E.~B.~Kolomeisky and J.~P.~Straley, 
Universality classes for line-depinning transitions, 
{\it Phys.~Rev.~B \bf 46} (1992), 12664--12674. 
 
\bibitem{L91} 
R.~Lipowsky, 
Typical and exceptional shape fluctuations of interacting strings, 
{\it Europhys.~Lett. \bf 15} (1991) 703--708. 
 
\bibitem{MBM01} 
D.~Marenduzzo, A.~Trovato and A.~Maritan, 
Phase diagram of force-induced DNA unzipping in exactly solvable models, 
{\it Phys.~Rev.~E \bf 64} (2001), 031901--031912; {\it cond-mat/0101207}. 
 
\bibitem{MBM02} 
D.~Marenduzzo, S.~M.~Bhattacharjee, A.~Maritan, E.~Orlandini and F.~Seno, 
Dynamical scaling of the DNA unzipping transition, 
{\it Phys.~Rev.~Lett. \bf 88} (2002), 28102--28105; {\it cond-mat/0103142}. 
 
\bibitem{MS93} 
N.~Madras and G.~Slade, 
{\it The Self-Avoiding Walk}, 
Birkh\"auser, Boston (1993). 
 
\bibitem{PB89} 
M.~Peyrard and A.~R.~Bishop, 
Statistical mechanics of a nonlinear model for DNA denaturation, 
{\it Phys.~Rev.~Lett. \bf 62} (1989) 2755--2758.  
 
\bibitem{PS66a} 
D.~Poland and H.~A.~Scheraga, 
Phase transitions in one dimension and the helix-coil transition in 
polyamino acids, 
{\it J.~Chem.~Phys. \bf 45} (1966), 1456--1463. 
 
\bibitem{PS66b} 
D.~Poland and H.~A.~Scheraga, 
Occurrence of a phase transition in nucleic acid models, 
{\it J.~Chem.~Phys. \bf 45} (1966), 1464--1469. 
 
\bibitem{PB95} 
T.~Prellberg and R.~Brak, 
Critical exponents from nonlinear functional equations for partially  
directed cluster models, 
{\it J.~Stat.~Phys. \bf 78} (1995), 701--730. 

\bibitem{SW88} 
D.~W.~Sumners and S.~G.~Whittington,
Knots in self-avoiding walks,
{\it J. Phys. A \bf 21} (1988) 1689--1694. 
 
\bibitem{TDP00} 
N.~Theodorakopoulos, T.~Dauxois and M.~Peyrard, 
Order of the phase transition in models of DNA thermal denaturation, 
{\it Phys.~Rev.~Lett. \bf 85} (2000), 6--9; {\it cond-mat/0004487}. 
 
\bibitem{WB85} 
R.~M.~Wartell and A.~S.~Benight, 
Thermal denaturation of DNA molecules: A comparison of theory with 
experiment, 
{\em Phys.~Rep. \bf126} (1985) 67--107. 
 
\end{thebibliography}
\end{document}